\documentclass[prl,twocolumn,showpacs,amsfonts,amsmath,floatfix]{revtex4}
\usepackage{graphicx}
\usepackage{psfrag}
\newcommand{\Journal}[4]{#1 {\bf #2}, #3 (#4)}
\newcommand{\PR}{Phys. Rev.}
\newcommand{\PRL}{Phys. Rev. Lett.}
\newcommand{\PRA}{Phys. Rev. A}
\newcommand{\JMP}{J. Math. Phys.}
\newcommand{\Science}{Science}
\newcommand{\PLA}{Phys. Lett. A}
\begin{document}
\title {Ground and excited states of spinor Fermi gases in tight waveguides 
and the Lieb-Liniger-Heisenberg model}
\author{M. D. Girardeau}
\email{girardeau@optics.arizona.edu}
\affiliation{College of Optical Sciences, University of Arizona,
Tucson, AZ 85721, USA}
\date{\today}
\begin{abstract}
The ground and excited states of a one-dimensional (1D) spin-$\frac{1}{2}$  
Fermi gas (SFG) with both attractive zero-range odd-wave interactions and 
repulsive zero-range even-wave interactions are mapped exactly 
to a 1D Lieb-Liniger-Heisenberg (LLH) model with delta-function
repulsions depending on isotropic Heisenberg spin-spin interactions, such that
the complete SFG and LLH energy spectra are identical. The ground state in the
ferromagnetic phase is given exactly by the Lieb-Liniger (LL) Bethe ansatz, 
and that in the antiferromagnetic phase by a variational method combining
Bethe ansatz solutions of the LL and 1D Heisenberg models. There are excitation
branches corresponding to LL type I and II phonons and spin waves, the latter 
behaving quadratically for small wave number in the ferromagnetic phase and 
linearly in the antiferromagnetic phase.
\end{abstract}
\maketitle
If an ultracold atomic vapor is confined in a de Broglie wave guide with 
transverse trapping so tight and temperature so low that the transverse
vibrational excitation quantum $\hbar\omega$ is larger than available
longitudinal zero point and thermal energies, the effective dynamics becomes   
one-dimensional (1D) \cite{Ols98,PetShlWal00}, a regime currently under 
intense experimental
investigation \cite{Tol04Mor04,Par04Kin04}. Confinement-induced 1D 
Feshbach resonances (CIRs) reachable by tuning the 1D 
coupling constant via 3D Feshbach scattering resonances \cite{Rob01}
occur for both Bose \cite{Ols98,BerMooOls03} and spin-aligned Fermi 
gases \cite{GraBlu04}, allowing tuning of the effective 1D coupling
constants to any desired values. Near the CIRs the ground states have
strong short-range correlations not representable by effective field
theories, and such systems have recently become the subject of extensive 
theoretical and experimental investigations because of their novelty and the 
recent experimental realization \cite{Par04Kin04} of the 1D gas of impenetrable
point bosons, now known as the Tonks-Girardeau (TG) gas. The Fermi-Bose 
(FB) mapping method, introduced in 1960 \cite{Gir60,Gir65} to solve the TG
problem exactly, and recently 
\cite{CheShi98,GirOls03,GirOls04-1,GirNguOls04,GraBlu04} extended to
a mapping between a 1D Bose gas with delta-function 
interactions of any strength [Lieb-Liniger (LL) gas \cite{LieLin63}] and a 
spin-aligned Fermi gas, will be generalized here to the Fermi gas with
unconstrained spins.  

The exact ground state of the ``fermionic TG'' (FTG) gas, 
a spin-aligned 1D Fermi gas with infinitely strongly attractive zero-range 
odd-wave interactions, has been determined recently 
\cite{GraBlu04,GirOls03,GirOls04-1,GirNguOls04} by FB mapping to the ideal Bose
gas, and its highly-correlated ground state was shown to possess 
superconductive 
off-diagonal long-range order (ODLRO) \cite{GirMin06,MinGir06}; this system
models a magnetically trapped ultracold gas of spin-$\frac{1}{2}$ fermionic 
atoms with very strong odd-wave interactions induced by a p-wave Feshbach 
resonance. If optically trapped instead, such a spinor Fermi gas (SFG) has 
richer properties, since its spins are unconstrained. In such a case, with both
spatial even-wave interactions
associated with spin singlet scattering and odd-wave interactions associated
with spin triplet scattering, one can vary the ratio of the two 
coupling constants by Feshbach resonance tuning of the odd-wave one,
leading to a rich phase diagram of ground state spin, which we have
determined recently \cite{Note1}. However, exact ground and excited states 
in the antiferromagnetic region of the phase diagram are unknown except for 
limiting cases where one of these interactions is negligible.
The purpose of this paper is to study the more general problem by a
generalized space-spin FB mapping to a 1D Lieb-Liniger-Heisenberg (LLH) model 
of particles with delta-function repulsions depending on Heisenberg 
spin-spin interactions, followed by a variational treatment combining 
Bethe ansatz solutions of the LL and 1D Heisenberg models. 

{\it Even and odd-wave SFG interactions:} The SFG is an ultracold vapor of 
fermionic atoms in 1D in two different hyperfine states, representable as 
spin-$\frac{1}{2}$ fermions. 
One can represent the $N$-atom states by space-spin wave functions 
$\psi_F(x_1,\sigma_1;\cdots;x_N,\sigma_N)$ where the $x_j$ are 1D spatial
coordinates and the spin z-component variables take on values 
$\pm\frac{1}{2}$ labelled by $\uparrow$ and $\downarrow$. This system is
assumed to be optically trapped on a ring of circumference $L$, so the $x_j$ 
are measured circumferentially and periodic boundary conditions 
with periodicity length $L$ are imposed. The Hamiltonian with LL units
$\hbar=2m=1$ in a limit of zero-range interactions is
\cite{GirOls03,GirOls04-1,GirNguOls04,Note1}
\begin{equation}\label{Fermi Hamiltonian}
\hat{H}_F=-\sum_{j=1}^{N}\partial_{x_j}^{2}
+\sum_{1\le j<\ell\le N}[g_{1D}^{e}\delta(x_{j\ell})\hat{P}_{j\ell}^s
+v_{1D}^{\text{o}}(x_{j\ell})\hat{P}_{j\ell}^t]\ .
\end{equation}
Here $x_{j\ell}=x_j-x_{\ell}$, $\delta$ is the Dirac delta function 
\cite{LieLin63}, $\hat{P}_{j\ell}^s$ ($\hat{P}_{j\ell}^t$) is the 
projector onto the subspaces of singlet (triplet) functions of the spin 
arguments $(\sigma_j,\sigma_\ell)$ for fixed values of all other arguments,  
and $v_{1D}^{\text{o}}$ is a strong, attractive, odd-wave 1D interaction of 
range $x_0\to 0+$ (1D analog of 3D p-wave interactions) occuring near the CIR
\cite{GraBlu04,GirOls03,GirOls04-1,GirNguOls04}. It is the zero-range limit of 
a deep and narrow square well of depth $V_0$ and width $2x_0$ with
$x_0\to 0+$ and $V_0\to\infty$ such that $V_0 x_0^2$ approaches a finite
limit, implying an attractive interaction stronger than a delta function, 
which induces a wave-function discontinuity in the zero-range limit 
\cite{CheShi98}. Taking the limit such that 
$\sqrt{V_0/2}=(\pi/2x_{0})[1+(2/\pi)^2 (x_{0}/a_{1D}^{o})]$
generates a two-body wave function satisfying the contact condition
$\psi_{F}(x_{j\ell}=0+)=-\psi_{F}(x_{j\ell}=0-)
= -a_{1D}^{o}\psi_{F}^{'}(x_{j\ell}=0\pm)$
where $a_{1D}^{o}<0$ is the 1D odd-wave scattering length and the prime denotes
differentiation \cite{GraBlu04,GirOls03,GirOls04-1,GirNguOls04}.
The even-wave 1D coupling constant $g_{1D}^{e}$ in (\ref{Fermi Hamiltonian})
is related to the even-wave scattering length $a_{1D}^{e}<0$ derived 
\cite{Ols98} from 3D s-wave scattering by $g_{1D}^{e}=-4/a_{1D}^e$ 
and the even-wave contact condition is the usual LL one \cite{LieLin63} 
$\psi_{F}'(x_{j\ell}=0+)
=-\psi_{F}'(x_{j\ell}=0-)=-(a_{1D}^{e})^{-1}\psi_{F}(x_{j\ell}=0\pm)$.

The $(j,\ell)$ singlet subspace is spanned by the bispinor wave function 
$\chi^s(\sigma_j,\sigma_{\ell})=\frac{1}{\sqrt{2}}
(\uparrow_j\downarrow_{\ell}-\downarrow_j\uparrow_{\ell})$ and the triplet
subspace by $\chi^t_1=\uparrow_j\uparrow_{\ell}$, $\chi^t_0=\frac{1}{\sqrt{2}}
(\uparrow_j\downarrow_{\ell}+\downarrow_j\uparrow_{\ell})$, and
$\chi^t_{-1}=\downarrow_j\downarrow_{\ell}$. Since $\psi_F$ is antisymmetric
under combined space-spin exchanges 
$(x_j,\sigma_j)\leftrightarrow(x_{\ell},\sigma_{\ell})$, the singlet is
the part of $\psi_F$ which is even in $x_{j\ell}$ as
$x_{j\ell}\to 0$ and the triplet is the part which is 
odd. In \cite{GirOls03,GirOls04-1,GirNguOls04,Note1} we accounted for this 
entirely by definition of the even and odd-wave pseudopotentials derived from 
3D s-wave and p-wave scattering, and omitted the
singlet and triplet projectors. However, here we study the SFG by
a mapping which converts the odd waves into even ones while leaving the
even ones unchanged, so it is essential to adjoin these projectors
so as not to lose track of this association. They
are related to the isotropic Heisenberg spin-spin interaction
$\hat{\mathbf{S}}_j\cdot\hat{\mathbf{S}}_{\ell}$ by
$\hat{P}_{j\ell}^s=\frac{1}{4}-\hat{\mathbf{S}}_j\cdot\hat{\mathbf{S}}_{\ell}$ 
and 
$\hat{P}_{j\ell}^t=\frac{3}{4}+\hat{\mathbf{S}}_j\cdot\hat{\mathbf{S}}_{\ell}$
where $\hat{\mathbf{S}}_j$ are the usual spin vector operators. 

{\it SFG-LLH mapping:} Assume that the system is contained in a 1D
ring trap of circumference $L$, with periodic
boundary conditions of periodicity length $L$.  
The spinor Fermi gas wave functions and Hamiltonian
can be mapped to those of a Lieb-Liniger-Heisenberg model, with wave
functions $\psi_{LLH}(x_1,\sigma_1;\cdots;x_N,\sigma_N)$ invariant under 
cyclic permutations 
$(x_1,\sigma_1;x_2,\sigma_2;\cdots;x_N,\sigma_N)\to
(x_2,\sigma_2;\cdots;x_N,\sigma_N;x_1,\sigma_1)$ around the ring, 
by multiplying the part of the SFG wave function $\psi_F$ which is odd in 
$x_{j\ell}$ as $x_{j\ell}\to 0$ by $\epsilon(x_{j\ell})$
where $\epsilon(x)=+1\ (-1)$ for $x>0\ (x<0)$, and $\epsilon(0)=0$, while
leaving the even part unchanged; this generalizes the original FB mapping of 
\cite{Gir60}. The $N$-particle configuration space decomposes into $N!$
permutation sectors, each defined by one particular ordering
of numerical values of $x_{1},\cdots,x_{N}$. In analogy with the LL model 
\cite{LieLin63} it is sufficient to define our spinor fermion wave functions 
$\psi_F$ only in the fundamental permutation sector $\mathfrak{R}_1$ wherein
$0\le x_{1}<x_{2}<\cdots<x_{N}\le N$ and to specify boundary conditions
on the boundary of $\mathfrak{R}_1$, which is an $(N-1)$-dimensional
hypersurface $\mathfrak{S}_1$. This boundary can be crossed in each of
$N$ orthogonal directions, each defined by a particular coordinate $x_j$
crossing its right-hand neighbor $x_{j+1}$ so that one enters a neighboring
permutation sector $\mathfrak{R}_{1,j}$ wherein
$x_{1}<x_{2}<\cdots<x_{j-1}<x_{j+1}<x_j<x_{j+2}<\cdots<x_{N}$. One obtains a 
well-posed problem for determination of the energy 
eigenfunctions $\psi_{F\alpha}$ and eigenvalues $E_{\alpha}$ by requiring that 
$\psi_{F\alpha}$ satisfy the free-particle Schr\"{o}dinger equation in the 
interior of $\mathfrak{R}_1$ plus  boundary conditions relating its value and 
normal gradient on $\mathfrak{S}_1$; $\psi_{F\alpha}$ can be uniquely 
extended to all $N!$ permutation sectors by space-spin antisymmetry. 

Since the spatially odd (even) part of $\psi_{F\alpha}$ is associated with 
the spin triplet (singlet) subspace, the desired  mapping 
$\psi_F\to\psi_{LLH}=\hat{U}\psi_F$ can be implemented by
a unitary mapping operator 
$\hat{U}=\hat{U}_N\hat{U}_{N-1}\cdots\hat{U}_1$ where each factor 
for $1\le j\le N-1$ is 
$\hat{U}_j=\hat{P}_{j,j+1}^s+\epsilon(x_{j+1}-x_j)\hat{P}_{j,j+1}^t$. 
$\hat{U}_j$ is unitary and self-inverse and 
changes the component of $\psi_F$ odd in $x_{j,j+1}$ around $x_{j,j+1}=0$ into 
an even one while leaving the even component unchanged. $\hat{U}_N$
is a special case required to close the ring periodically:
$\hat{U}_N=\hat{P}_{N,1}^s+\epsilon(x_1+L-x_N)\hat{P}_{N,1}^t$. 
Since $x_j<x_{j+1}$ in $\mathfrak{R}_1$, the $\epsilon(x_{j+1}-x_j)$ factors 
are all unity there, and in $\mathfrak{R}_1$ the mapping reduces to the 
identity. The term $\epsilon(x_{j+1}-x_j)$ in $\hat{U}_j$ is to be interpreted
as a prescription for extension to the adjacent permutation sector 
$\mathfrak{R}_{1,j}$. In 
$\mathfrak{R}_{1,j}$ all factors $\hat{U}_{\ell}$ with $\ell\ne j$ are
unity, verifying that $\hat{U}$ maps $\psi_F$ to a LLH wave function 
$\psi_{LLH}$ differing from $\psi_F$ only by extension from $\mathfrak{R}_1$ 
into adjacent sectors $\mathfrak{R}_{1,j}$ by multiplication of the component 
of $\psi_F$ odd in $x_{j,j+1}$ around $x_{j,j+1}=0$ by $\epsilon(x_{j+1}-x_j)$.

When extended to all permutation sectors  
$\psi_{LLH}$ is neither bosonic nor fermionic under combined space-spin
exchanges $(x_j,\sigma_j)\leftrightarrow(x_{\ell},\sigma_{\ell})$; 
it has both Bose-like components which are space-even and spin-even,
and Fermi-like components which are space-even and spin-odd, but no
space-odd components. 
The mapping converts the odd-spatial-wave, spin triplet contact 
discontinuities of $\psi_F$ into LL contact cusps where $\psi_F$ is
continuous and its derivative changes sign, while retaining the original LL
even-spatial-wave, spin singlet cusps. As in 
\cite{GirOls03,GirOls04-1,GirNguOls04} this changes the odd spatial wave, 
spin triplet contact condition 
$\psi_{F}(x_{j\ell}=0+)=-\psi_{F}(x_{j\ell}=0-)
= -a_{1D}^{o}\psi_{F}^{'}(x_{j\ell}=0\pm)$ 
of the SFG into an even spatial wave, spin triplet contact condition 
$\psi_{LLH}'(x_{j\ell}=0+)
=-\psi_{LLH}'(x_{j\ell}=0-)=-(a_{1D}^{o})^{-1}\psi_{LLH}(x_{j\ell}=0\pm)$
for the LLH gas with the same scattering length $a_{1D}^o$. 
Then both singlet and triplet contact conditions are of the usual form 
generated by LL delta functions \cite{LieLin63} and it follows 
that $\psi_{LLH\alpha}$ is an eigenstate of $\hat{H}_{LLH}$ with energy 
$E_\alpha$ if $\psi_{F\alpha}$ is an eigenstate of $\hat{H}_F$ with energy 
$E_\alpha$ and conversely, where after substituting  
$\hat{P}_{j\ell}^s=\frac{1}{4}-\hat{\mathbf{S}}_j\cdot\hat{\mathbf{S}}_{\ell}$ 
and 
$\hat{P}_{j\ell}^t=\frac{3}{4}+\hat{\mathbf{S}}_j\cdot\hat{\mathbf{S}}_{\ell}$
one finds
\begin{flushleft}
\begin{equation}\label{LLH Hamiltonian}
\hat{H}_{LLH}=-\sum_{j=1}^{N}\partial_{x_j}^{2}
+\sum_{1\le j<\ell\le N}[\frac{3c+c'}{2}
+2(c-c')\hat{\mathbf{S}}_j\cdot\hat{\mathbf{S}}_{\ell}]\delta(x_{j\ell})]
\end{equation}
\end{flushleft}
with coupling constants given in terms of the odd and even-wave SFG
scattering lengths $a_{1D}^o$ and $a_{1D}^e$ by
$c=2/|a_{1D}^o|$ and $c'=\frac{1}{2}g_{1D}^e=2/|a_{1D}^e|$. 

{\it Ground state phase diagram:} Assume that $N$
is even. We previously determined \cite{Note1} 
the phase diagram of total ground-state spin $S$ in the $(c,c')$ plane,
finding (a) for $c<c'$ the ground state is ferromagnetic with $S=N/2$; 
(b) for $c>c'$ it is antiferromagnetic with $S=0$; (c) for $c=c'$ it is
highly degenerate, all (integer) values of $S$ from $0$ to $\frac{N}{2}$
and $S_z$ from $-\frac{N}{2}$ to $\frac{N}{2}$ having the same energy.
(c) now follows trivially from Eq. (\ref{LLH Hamiltonian}); 
if $c=c'$ the $\hat{\mathbf{S}}_j\cdot\hat{\mathbf{S}}_{\ell}$ terms 
in (\ref{LLH Hamiltonian}) vanish, so the entire energy spectrum is 
spin-independent.

\begin{figure}
 \psfrag{E0LL}{$E_{0}/n^2N$}
 \psfrag{c}{$c/n$}
\includegraphics[width=7.5cm,angle=0]{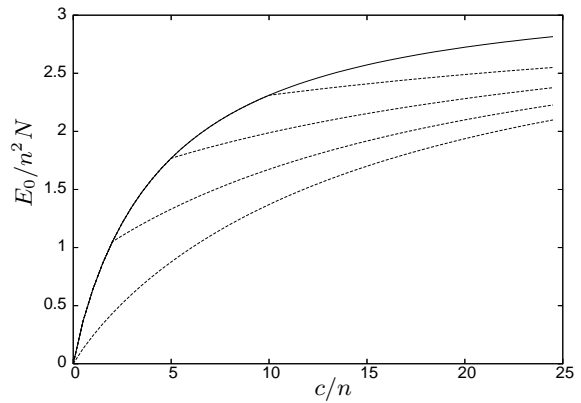} 
  \caption{Ground state energy of the 1D spinor Fermi gas as a 
function of $c=2/|a_{1D}^o|$ for fixed values of $c'=2/|a_{1D}^e|$, where
$a_{1D}^o$ and $a_{1D}^e$ are odd and even-wave 1D scattering 
lengths. Solid line: Exact ground state energy in the ferromagnetic phase,
stable only for $c<c'$. Dashed lines: Variational ground state energy in the
antiferromagnetic phase, stable only for $c>c'$, for 
$c'/n=0,\ 2,\ 5$, and $10$ (bottom to top).}  \label{fig1}
\end{figure}

{\it Ferromagnetic phase:} In this phase, which occurs for $c<c'$, the ground 
state is totally spin-aligned ($S=N/2$), there are no quantum fluctuations of 
$\hat{\mathbf{S}}_j\cdot\hat{\mathbf{S}}_{\ell}$ which may therefore be 
replaced by its expectation value $\frac{1}{4}$, and the square bracket in 
(\ref{LLH Hamiltonian}) reduces simply to $2c\delta(x_{j\ell})$, the usual
LL interaction, independent of $c'$. Then the LL Bethe ansatz \cite{LieLin63}
yields the exact ground state for all $c$, as in 
\cite{GraBlu04,GirOls03,GirOls04-1,GirNguOls04}. As $c$ increases from zero 
for fixed $c'$, the system remains in the ferromagnetic phase until
$c$ reaches $c'$, where it undergoes a quantum phase transition to the
$S=0$ antiferromagnetic phase. This is illustrated in Fig.~\ref{fig1},
where the solid line shows the ground state energy $E_{0}(c)$
in the ferromagnetic phase \cite{LieLin63,Note3}  
and the dashed lines for fixed values of $c'$ show the energy in the 
antiferromagnetic phase. 

{\it Antiferromagnetic phase:} Now assume $c>c'$, in which case the ground
state is highly nontrivial, has $S=0$, and depends on both $c$ and $c'$.
As before, it is sufficient to restrict ourselves to the sector 
$\mathfrak{R}_1$: $0<x_1<x_2<\cdots<x_N<L$. One can obtain an upper bound
on the ground state energy by taking a variational state which is a product of
a purely spatial state $\psi_{\text{space}}(x_1,\cdots,x_N)$ by a purely spin
state $\psi_{\text{spin}}(\sigma_1,\cdots,\sigma_N)$. In spite of its product
structure, this trial state incorporates some space-spin correlation implicitly
via the projectors 
$\hat{P}_{j\ell}^s=\frac{1}{4}-\hat{\mathbf{S}}_j\cdot\hat{\mathbf{S}}_{\ell}$ 
and 
$\hat{P}_{j\ell}^t=\frac{3}{4}+\hat{\mathbf{S}}_j\cdot\hat{\mathbf{S}}_{\ell}$.
Taking $\psi_{\text{spin}}$ to be the ground state of the isotropic, 
antiferromagnetic Heisenberg Hamiltonian
$\hat{H}_{\text{Heis}}=\sum_{j=1}^{N-1} 
\hat{\mathbf{S}}_j\cdot\hat{\mathbf{S}}_{j+1}
+\hat{\mathbf{S}}_N\cdot\hat{\mathbf{S}}_1$, for which in the thermodynamic
limit 
$E_{0,\text{Heis}}/N=\frac{1}{4}-\ln 2$ \cite{Hul38}, reduces the Hamiltonian
to an LL model \cite{LieLin63,Note2}, yielding an upper bound
$E_0(c,c')<E_{0,LL}(c_{LL})$ with $c_{LL}=c(1-\ln 2)+c'\ln 2$. The product 
state $\psi_{\text{space}}\psi_{\text{spin}}$ is exact in the ferromagnetic 
phase (where $\hat{\mathbf{S}}_j\cdot\hat{\mathbf{S}}_{j+1}=\frac{1}{4}$), on 
the phase boundary $c=c'$, and when $c=0$ \cite{Gau67,Yan67}, so it may be a 
good approximation throughout the $(c,c')$ plane. This variational ground
state energy, obtained from the LL ground-state energy  
\cite{Note3}, is shown by the dashed lines in  Fig.~\ref{fig1}.

{\it Excitations:} Both phonons (number density waves) and spin waves (spin 
density waves) occur in this system. Since a space-spin product variational
approximation was used for the ground state, the energies of these
excitations will be evaluated using variational states 
$\psi_{\text{space}}\psi_{\text{spin}}$, but now with either 
$\psi_{\text{space}}$ or $\psi_{\text{spin}}$ excited. For one-phonon 
excitations $\psi_{\text{space}}$ is an exact one-phonon LL 
excited state \cite{Lie63} and $\psi_{\text{spin}}$ is the exact
ground state of the 1D Heisenberg model. Consider first such excitations 
from the ferromagnetic phase ($c<c'$). There are two such modes: Hole
excitations (Lieb's type II excitations \cite{Lie63}) correspond to
promotion of a particle from quasimomentum $K-k$ inside the LL-Fermi
sea (Fig. 2 of LL \cite{LieLin63}) to just above the top 
(quasimomentum $K$) of the sea, whereas umklapp excitations (Lieb's
type I excitations) correspond to promotion of a particle from $-K$ to
$K+k$. Both have the same initial slope corresponding to a sound speed
given, assuming LL units $\hbar=2m=1$, by Lieb's Eqs. (1.4) and (1.5), 
$v_s=2\sqrt{\mu-\frac{1}{2}\gamma\dot{\mu}}$. Here $\mu$ is the LL ground
state chemical potential $\mu=n^2(3e-\gamma\dot{e})$, the dot denotes
differentiation with respect to $\gamma$, $n$ is the 1D atom number
density, $e(\gamma)=n^{-2}(E_0/N)$, $E_0$ is the LL ground state energy,
and $\gamma=c/n$ where $c$ is the coupling constant in the LL interaction
$2c\delta(x_{j\ell})$ \cite{Lie63}. The type I and II  
curves as a function of momentum are shown in Lieb's Figs. 1-3. The
corresponding phonon excitations from the antiferromagnetic phase ($c>c'$)
differ only by replacement of $c$ by $c_{LL}=c(1-\ln 2)+c'\ln 2$.   

\begin{figure}
 \psfrag{Scaled spin wave energy}{$\text{Scaled spin wave energy}$}
 \psfrag{q}{$\text{q}$}
\includegraphics[width=7.5cm,angle=0]{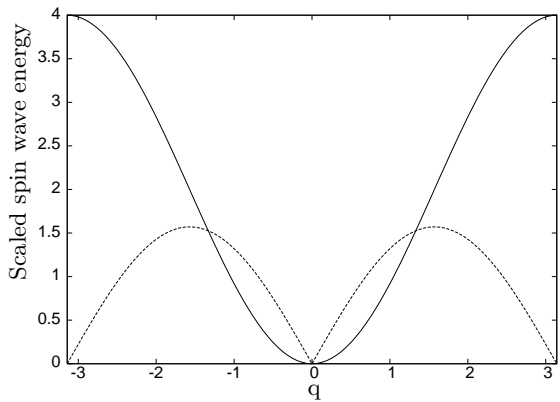} 
 \caption{Solid line: $\epsilon_{\text{ferro}}(q)/(c'-c)n^3 e'(c/n)$. 
Dashed line: $\epsilon_{\text{antiferro}}(q)/(c-c')n^3 e'(c_{LL}/n)$.}
  \label{fig2}
\end{figure}

Determination of spin wave energies is more delicate.
The total energy is the expectation value of 
(\ref{LLH Hamiltonian}) in a space-spin product state 
$\psi_{\text{space}}\psi_{\text{spin}}$ in which $\psi_{\text{space}}$ is
the ground state and $\psi_{\text{spin}}$ is a one spin wave excited
state. The spin wave
energy enters only indirectly via the expectation values
$\langle\hat{\mathbf{S}}_j\cdot\hat{\mathbf{S}}_{j+1}\rangle$ which affect the 
LL coupling constant 
$c_{LL}=\frac{3c+c'}{4}+(c-c')\langle\hat{S}_j\cdot\hat{S}_{j+1}\rangle$.
Consider first spin-wave excitations in the ferromagnetic phase ($c<c'$),
for which $\langle\hat{S}_j\cdot\hat{S}_{j+1}\rangle_0=\frac{1}{4}$ for
all $j$, and $c_{LL}=c$. In the excited state  
it is still independent of $j$, but now 
$\langle\hat{S}_j\cdot\hat{S}_{j+1}\rangle
=\langle\hat{S}_j\cdot\hat{S}_{j+1}\rangle_0+\Delta(q)$ where $\Delta(q)$ is 
the change due to the presence of a spin wave of wave number $q$, 
given by $\Delta(q)=-\epsilon_{\text{Heis}}(q)/N$ \cite{Note4} with 
$\epsilon_{\text{Heis}}(q)$ the Heisenberg spin wave energy. For 
$\hat{H}_{\text{Heis}}=-\sum_j \hat{S}_j\cdot\hat{S}_{j+1}$ one has
\cite{LieMat66} $\epsilon_{\text{Heis}}(q)=2(1-\cos q)$ and
$\Delta(q)=-2N^{-1}(1-\cos q)$, yielding a change
$-2N^{-1}(c-c')(1-\cos q)$ in the effective LL coupling constant from the
expectation value of Eq. (\ref{LLH Hamiltonian}). Then Taylor expansion
about the ground state energy of $\hat{H}_{LLH}$ yields the SFG spin wave
energy $\epsilon_{\text{ferro}}(q)=2(c'-c)(1-\cos q)n^3 e'(\gamma)$ where 
$e'(\gamma)$ is the derivative of $e(\gamma)=E_0/n^{2}N$, $\gamma=c/n$, and 
$n$ is the 1D number density \cite{LieLin63,Note3}. 
$\epsilon_{\text{ferro}}(q)$ is quadratic for $q$, as is also the case for
isospin waves in a 1D two-component Bose gas \cite{Li03,Fuc05}. 
The antiferromagnetic case is almost the same, except that the change
in $\langle\hat{\mathbf{S}}_j\cdot\hat{\mathbf{S}}_{j+1}\rangle$ from
its ground state value $\frac{1}{4}-\ln 2$ is now positive and derived
from the antiferromagnetic spin wave energy \cite{CloPea62}
$\epsilon_{\text{Heis}}(q)=\frac{\pi}{2}|\sin q|$, yielding a SFG spin 
wave energy 
$\epsilon_{\text{antiferro}}(q)=\frac{\pi}{2}(c-c')|\sin q|n^3 e'(\gamma_{LL})$
with $\gamma_{LL}=c_{LL}/n$ and $c_{LL}=c(1-\ln 2)+c'\ln 2$. This is linear 
for small $q$. These spin wave energies are shown in Fig.~\ref{fig2};
the values of $c$ and $c'$ enter only in scale factors. 
%

{\it Conclusions:} Ground and low excitation energies of the 1D spinor
Fermi gas with both even and odd-wave scattering have been determined by
an exact mapping to a ``Lieb-Liniger-Heisenberg'' model which was treated
by a variational method combining Bethe ansatz solutions of the LL
and Heisenberg models. The richness of ground state phases
and excitations suggests that experimental study of this system could be quite
fruitful.
\begin{acknowledgments}
This research was supported by U.S. Office of Naval Research grants 
N00014-03-1-0427 and N00014-06-1-0455 through subcontracts from the University 
of Southern California. I thank Maxim Olshanii for helpful comments, and 
Anna Minguzzi for suggestions and for help with Fig. 1.

\end{acknowledgments}

\begin{thebibliography}{37}
%
\bibitem{Ols98} M. Olshanii, \Journal{\PRL}{81}{938}{1998}.
%
\bibitem{PetShlWal00} D.S. Petrov, G.V. Shlyapnikov, and J.T.M. Walraven,
\Journal{\PRL}{85}{3745}{2000}.
%
\bibitem{Tol04Mor04} B.L. Tolra {\it et al.}, \Journal{\PRL}{92}{190401}{2004};
T. St\"{o}ferle , H. Moritz, C. Schori, M. K\"{o}hl, 
and T. Esslinger, {\it ibid.} {\bf 92}, 130403 (2004).
%
\bibitem {Par04Kin04} B. Paredes, {\it et al.},
Nature {\bf 429}, 277 (2004);
 T. Kinoshita, T. Wenger, and D.S. Weiss, \Journal{\Science}
{305}{1125}{2004}.
%
\bibitem{Rob01} J.L. Roberts {\it et al.}, \Journal{\PRL}{86}{4211}{2001}.
%
\bibitem{BerMooOls03} T. Bergeman, M. Moore, and M. Olshanii,
\Journal{\PRL}{91}{163201}{2003}.
%
\bibitem{GraBlu04} B.E. Granger and D. Blume, \Journal{\PRL}{92}{133202}{2004}.
%
\bibitem{Gir60} M. Girardeau, \Journal{\JMP}{1}{516}{1960}.
%
\bibitem{Gir65} M.D. Girardeau, \Journal{\PR}{139}{B500}{1965}, Secs. 2, 3, 
and 6.
%
\bibitem{CheShi98} T. Cheon and T. Shigehara, \Journal{\PLA}{243}{111}{1998}
and \Journal{\PRL}{82}{2536}{1999}.
%
\bibitem{GirOls03} M.D. Girardeau and M. Olshanii, cond-mat/0309396.
%
\bibitem{GirOls04-1} M.D. Girardeau and M. Olshanii, 
\Journal{\PRA}{70}{023608}{2004}.
%
\bibitem{GirNguOls04} M.D. Girardeau, Hieu Nguyen, and M. Olshanii,
Optics Communications {\bf 243}, 3 (2004).
%
\bibitem{LieLin63} Elliott H. Lieb and Werner Liniger, Phys. Rev. {\bf 130},
1605 (1963).
%
\bibitem{GirMin06} M.D. Girardeau and A. Minguzzi, 
\Journal{\PRL}{96}{080404}{2006}.
%
\bibitem{MinGir06} A. Minguzzi and M.D. Girardeau, 
\Journal{\PRA}{73}{063614}{2006}.
%
\bibitem{Note1} See Sec. 3.3 of \cite{GirNguOls04}, particularly pp. 19 and 
20. 
%
\bibitem{Note3} The LL ground state energy and its first derivative 
are made available online by V. Dunjko and M. Olshanii at 
http://physics.usc.edu/$\widetilde{\hspace{0.15cm}}$olshanii/DIST/.
%
\bibitem{Li03} Y.-Q. Li, S.-J. Gu, Z.-J. Ying, and U. Eckern,
Europhys. Lett. {\bf 61}, 368 (2003).
%
\bibitem{Fuc05} J.N. Fuchs, D.M. Gangardt, T. Keilmann, and G.V. Shlyapnikov,
\Journal{\PRL}{95}{150402}{2005}.
%
\bibitem{Hul38} L. Hulth\'{e}n, Arkiv. Math. Astron. Fys. {\bf 26A},  No. 11 
(1938).
%
\bibitem{Note2} Since the ground state is a momentum eigenstate (with
eigenvalue zero), 
$\langle\hat{\mathbf{S}}_j\cdot\hat{\mathbf{S}}_{j+1}\rangle$ 
is independent of $j$, so in the Heisenberg ground state 
$\langle\hat{\mathbf{S}}_j\cdot\hat{\mathbf{S}}_{j+1}\rangle>=E_{0,\text{Heis}}/N$.
%
\bibitem{Gau67} M. Gaudin, Phys. Lett. {\bf 24A}, 55 (1967).
%
\bibitem{Yan67} C.N. Yang, \Journal{\PRL}{19}{1312}{1967}.
%
\bibitem{Lie63} Elliott H. Lieb, Phys. Rev. {\bf 130}, 1616 (1963).
%
\bibitem{Note4} The minus sign is because the spin wave \emph{decreases}
$\langle\hat{\mathbf{S}}_j\cdot\hat{\mathbf{S}}_{j+1}\rangle$ from its
maximal value $\frac{1}{4}$. The allowed values of $q$ are 
$q=\nu 2\pi/N$ with $\nu=0,\pm 1,\pm 2,\cdots$.
%
\bibitem{LieMat66} Elliott H. Lieb and Daniel C. Mattis, 
\textit{Mathematical Physics in One Dimension} (Academic Press, New York 1966),
p. 459.
%
\bibitem{CloPea62} J. des Cloizeaux and J.J. Pearson, 
\Journal{\PR}{128}{2131}{1962}.
%
\end{thebibliography}
\end{document}